\documentclass[twocolumn,english,prb,showpacs]{revtex4-1}
\usepackage[colorlinks=true,urlcolor=blue,citecolor=blue,linkcolor=blue]{hyperref}

\usepackage[T1]{fontenc}
\usepackage[latin9]{inputenc}
\usepackage{amssymb}
\usepackage{graphicx}
\usepackage{amsmath,color}
\usepackage{mathrsfs}
\usepackage{float}
\usepackage{indentfirst}
\usepackage{subfigure}
\usepackage{multirow}
\usepackage{booktabs}
\newcommand{\ra}[1]{\renewcommand{\arraystretch}{#1}}
\DeclareMathOperator{\Tr}{Tr}
\DeclareMathOperator{\diag}{diag}

\makeatletter

\begin{document}

\title{Efficient Continuous-time Quantum Monte Carlo Method for\\ the Ground State of Correlated Fermions}

\author{Lei Wang$^{1}$, Mauro Iazzi$^{1}$, Philippe Corboz$^{2}$ and Matthias Troyer$^{1}$}
\address{$^{1}$Theoretische Physik, ETH Zurich, 8093 Zurich, Switzerland}
\address{$^{2}$Institute for Theoretical Physics, University of Amsterdam, Science Park 904 Postbus 94485, 1090 GL Amsterdam, The Netherlands}

\begin{abstract}
We present the ground state extension of the efficient quantum Monte Carlo algorithm for lattice fermions of arXiv:1411.0683. Based on continuous-time expansion of imaginary-time projection operator, the algorithm is free of systematic error and scales \emph{linearly} with projection time and interaction strength. Compared to the conventional quantum Monte Carlo methods for lattice fermions, this approach has greater flexibility and is easier to combine with powerful machinery such as histogram reweighting and extended ensemble simulation techniques. We discuss the implementation of the continuous-time projection in detail using the spinless $t-V$ model as an example and compare the numerical results with exact diagonalization, density-matrix-renormalization-group and infinite projected entangled-pair states calculations. Finally we use the method to study the fermionic quantum critical point of spinless fermions on a honeycomb lattice and confirm previous results concerning its critical exponents.   
\end{abstract}


\maketitle

\section{Introduction}
%

Quantum Monte Carlo (QMC) methods are powerful and versatile  tools for studying quantum phases and phase transitions. Algorithmic development in past two decades including the nonlocal updates~\cite{Evertz:1993cm,Evertz:2003ch,Prokofev:1998tc, Sandvik:1999wh, Syljuasen:2002hw} and the continuous-time formulations~\cite{Nikolai96,PhysRevLett.77.5130} have greatly boosted the power of QMC methods, even surpassing the hardware improvements following Moore's law. Using the modern QMC methods, the simulation of bosons and unfrustrated spin models is considered as a solved problem. QMC simulations therefore can be used to test novel theoretical scenarios~\cite{Senthil:2004cm, Sandvik:2007dt, Balents:2002kg, Isakov:2012ds, Kaul:2013ika} and to verify experimental realizations.~\cite{Trotzky:2010gh}

While efficient algorithms exist for the simulation of bosons and unfrustrated spin models,~\cite{Evertz:1993cm,Evertz:2003ch,Prokofev:1998tc, Sandvik:1999wh, Syljuasen:2002hw,Troyer:2003fta,Alet:2005cv} simulations of fermions are more challenging than bosons and quantum spins because of the infamous fermionic sign problem.~\cite{LohJr:1990up, Troyer:2005hv} It causes exponential growth of computational effort as system size or inverse temperature increases. Even for systems without a sign problem, the phase diagram of correlated fermions can be nontrivial to establish,~\cite{Meng:2010gc,Sorella:2012hia} not to mention to accurately determine the universality class and associated critical exponents.~\cite{Assaad:2013kg, PhysRevB.89.205403} The main reason for this difficulty is the unfavorable \emph{superlinear} scaling with system size and/or inverse temperature of determinantal quantum Monte Carlo methods, which are the workhorse of correlated lattice fermion simulations. 

Determinantal QMC method sums a factorially large number of fermion exchange processes into a matrix determinant, thereby  avoiding the fermion sign problems in certain cases. The first algorithm based on this idea is the Blankenbecler-Scalapino-Sugar (BSS) method.~\cite{Blankenbecler:1981vj} It maps an interacting fermionic system to free fermions in a spatially and temporally fluctuating external field and then performs Monte Carlo sampling of this field.  Numerical instabilities of the original approach have been remedied in Refs.~\onlinecite{White:1989wh, Loh:1992}. The BSS algorithm has become the method of choice of many lattice fermion simulations due to its linear scaling in the inverse temperature $\beta$.  We refer to Refs.~\onlinecite{DosSantos:610304,Assaad:2008hx} for pedagogical reviews.

Closely related  is the Hirsch-Fye algorithm,~\cite{PhysRevLett.56.2521} which is numerically more stable and is more broadly applicable because it is formulated using a (potentially time-dependent) action rather than a Hamiltonian. However, its computational effort scales \emph{cubically} with the inverse temperature and the interaction strength therefore is much less efficient than the BSS method for the cases where both methods are applicable. The Hirsch-Fye method thus has typically been used in the study of quantum impurity problems and as impurity solvers in the framework of  dynamical mean field theory (DMFT),~\cite{Anonymous:z_AfEOwS} where time-dependent actions need to be simulated.

\begin{table*}[!htbp]
\caption{Comparison between various determinantal QMC methods for fermions. The ground state methods are extensions of the corresponding finite temperature methods. They have similar scalings when replace the inverse temperature $\beta$ by the projection time $\Theta$. $N$ denotes the number of correlated sites and $V$ denotes the interaction strength.~\label{tab:comparison}}
\ra{1.5}
\begin{tabular}{lcccc cccc}\toprule
 & \multicolumn{4}{c}{Lattice Models} & \multicolumn{4}{c}{Impurity Models} \\
   \cmidrule(rr){2-5} \cmidrule(rr){6-9}
Method name&  BSS  &  --- & LCT-INT & LCT-AUX  & Hirsch-Fye  & CT-INT & CT-AUX  & CT-HYB\\\midrule
Finite temperature & Ref.~\onlinecite{Blankenbecler:1981vj} & \multirow{2}{*}{ Ref.~\onlinecite{PhysRevLett.82.4155}}  & \multicolumn{2}{c}{Ref.~\onlinecite{Mauro}}  & Ref.~\onlinecite{PhysRevLett.56.2521} & Ref.~\onlinecite{Rubtsov:2005iw} & Ref.~\onlinecite{Gull:2008cm} & Ref.~\onlinecite{PhysRevLett.97.076405}  \\   
Ground state &Ref.~\onlinecite{Sugiyama:1986vt, Sorella:1989uk, White:1989wh}  &  & This paper & --- & Ref.~\onlinecite{Feldbacher:2004gz}  & Ref.~\onlinecite{Assaad:2007be}& --- & --- \\  \midrule 
Trotter error & Yes  & No & No & No & Yes & No &No & No  \\ 
Auxiliary field  & Yes & Yes & No & Yes & Yes & No & Yes & No \\
Scaling  &  $\beta  V N^{3}$  \footnote{Although the number of operations does not \emph{explicitly} depend on the interaction strength $V$, one needs to increase the number of time slices proportional to $V$ to keep a constant coupling strength with the auxiliary field, i.e. to retain the same level of fluctuations.}  
& \footnote{The scaling of this code is unclear since it is not discussed in  Ref.~\onlinecite{PhysRevLett.82.4155} and important implementation details are missing.}  &  $\beta V N^{3}$ & $\beta V N^{3}$ & $(\beta V N)^{3}$ &$ (\beta V N)^{3}$ &$ (\beta V N)^{3}$  & $e^{N}$ \\
 \bottomrule 
\end{tabular}
\end{table*}

Both the BSS and the  Hirsch-Fye algorithm are based on a discretization of imaginary time, thus introducing a systematic time step error, called the Trotter error. Nearly twenty years ago it was realized that the time-discretization is not necessary for the simulation of lattice models.~\cite{Nikolai96,PhysRevLett.77.5130} Besides increased accuracy due to the absence of a Trotter error, continuous imaginary time formulations often results in a more efficient and flexible algorithm.~\cite{Prokofev:1998tc} In Ref.~\onlinecite{PhysRevLett.82.4155} a  first continuous-time QMC method for lattice fermions has been proposed. However the scaling of this algorithm and numerical stabilization have not been discussed in this paper and we are not aware of any application of the algorithm. Further development on fermionic continuous-time QMC algorithms \cite{Gull:2011jd} have focused on quantum impurity problems: the continuous-time interaction expansion (CT-INT) algorithm~\cite{Rubtsov:2005iw}, continuous-time hybridization expansion (CT-HYB)  algorithm\cite{PhysRevLett.97.076405} and the continuous-time auxiliary field (CT-AUX)~\cite{Gull:2008cm} algorithm. CT-INT and CT-AUX are based on weak-coupling expansion of the action and share the same scaling as the Hirsch-Fye method.~\cite{Mikelsons:2009eka} 
These methods have revolutionized the simulation of quantum impurity problems and DMFT calculations.~\cite{Gull:2011jd} However, for lattice models they remained suboptimal compared to the BSS method due to their cubic scaling in the inverse temperature. Very recently an efficient continuous-time algorithm has been developed by two of the authors that scales identically to the time-honored BSS method~\cite{Mauro} and can be used both with an auxiliary field (LCT-AUX) and without (LCT-INT). The prefix ``L'' indicating both their \emph{linear} scaling and their applicability to \emph{lattice} models. In Table~\ref{tab:comparison} we summarize some properties of these determinantal QMC methods.

Finite-temperature determinantal QMC methods can be extended to projector formulations,~\cite{Sugiyama:1986vt, Sorella:1989uk, White:1989wh, Feldbacher:2004gz, Assaad:2007be} where the ground state is obtained from imaginary time projection of a trial wave function. In addition to being more direct to address quantum phases at zero temperature, the projector formalism often allows for additional optimizations such as symmetry and quantum number projections~\cite{Capponi:2001jc, Shi:2014bw} and combinations with fixed-node ideas in the presence of a sign problem.~\cite{PhysRevLett.74.3652} In the case of the BSS method, numerical stabilization also becomes easier in the ground state formulation.~\cite{Loh:1992, Assaad:2008hx} On the other hand, for projection methods it is crucial to achieve a linear scaling in the projection time since the results are exact only in the limit of infinite projection time.~\footnote{In practice, the projection time should be longer than the inverse of the minimal energy gap of the system} The ground state variants of the Hirsch-Fye and the CT-INT methods~\cite{Feldbacher:2004gz,Assaad:2007be} exhibit \emph{cubic} scaling and thus are not ideal for lattice model simulations.

In this paper we present details of the projection version of the LCT-INT method whose feasibility has already been mentioned in Ref.~\onlinecite{Mauro}. This algorithm  provides an efficient continuous-time projection QMC approach for ground state simulations of correlated fermions. It retains the \emph{linear scaling} with projection time and matches the one of the widely applied projector BSS method~\cite{Sugiyama:1986vt, Sorella:1989uk, White:1989wh, Meng:2010gc,Sorella:2012hia, Assaad:2013kg} while completely eliminating the time discretization error. Moreover, the continuous-time formulation has greater flexibility for measuring observables and can easily be combined with histogram reweighting~\cite{1988PhRvL..61.2635F, Ferrenberg:1989tf} and extensive ensemble simulation~\cite{Wang:2001eba, Troyer:2003fta, Trebst:2014p18439} techniques.

The organization of this paper is as follows, in Section~\ref{sec:model} we introduce a model system of spinless fermions that we will use to explain the algorithm in Section~\ref{sec:algorithm}. Section~\ref{sec:results} contains comparisons of the method with other numerical approaches and results on the quantum critical point of spinless fermions on a honeycomb lattice. We end with discussions of future prospects in Sec.~\ref{sec:discussion}.

\section{Model \label{sec:model}}
To make the presentation of our algorithm more concrete, we will consider the following spinless fermion model at half-filling: 

\begin{eqnarray}
 \hat{H}& = & \hat{H}_{0} + \hat{H}_{1}, 
   \label{eq:Ham} \\
  \hat{H}_{0} & = & -t \sum_{\langle {i,j} \rangle} \left( \hat{c}_{{i}}^{\dagger} \hat{c}_{{j}} + \hat{c}_{{j}}^{\dagger} \hat{c}_{{i}}\right)
   \equiv \sum_{i,j} \hat{c}^{\dagger}_{{i}} K_{ij} \hat{c}_{{j}},  \label{eq:K} \\
  \hat{H}_{1} & = & V \sum_{\langle {i,j} \rangle} \left( \hat{n}_{{i}} - \frac{1}{2}
  \right) \left( \hat{n}_{{j}} - \frac{1}{2} \right), 
  \label{eq:V}
\end{eqnarray} 
where $\hat{c}_{i}$ is the fermion annihilation operator. $t$ denotes the nearest-neighbor tunneling, $V>0$ denotes the extended Hubbard repulsive interaction, and we have introduced the matrix $K$ to denote the single particle matrix elements.

Quantum Monte Carlo studies of this model on a square lattice date back to the early days of the BSS method.~\cite{Scalapino:1984wz, Gubernatis:1985wo} However, these simulations suffer from the fermion sign problem because the Monte Carlo weight is a \emph{single} determinant which is not guaranteed to be positive in general. Recently it was discovered that the model (\ref{eq:Ham}) is \emph{naturally} free from the sign problem on bipartite lattices at half-filling in the CT-INT formulation,\cite{Huffman:2014fj, Wang:2014ib} because the Monte Carlo weight can be expressed as the determinant of a \emph{real skew-symmetric} matrix. This determinant equals the square of the matrix Pfaffian and is thus nonnegative. A conventional auxiliary field decomposition, on the other hand, breaks this symmetry. It was shown that this model also allows sign problem free simulation in the BSS formalism if one works in a Majorana fermion representation,\cite{Li:2014tm,Li:2014vgb} i.e. performs the auxiliary field decomposition not in the density channel but in the pairing channel. The idea applies not only to the BSS algorithm but can be generalized to the continuous-time QMC algorithm.\cite{Mauro}

On the honeycomb lattice, this model exhibits a quantum phase transition from a Dirac semimetal to a charge-density-wave (CDW) phase. The quantum critical point is unconventional because of the coupling of CDW order parameter to the low-energy Dirac fermions.~\cite{Herbut:2006jaa, Herbut:2009bba}  Simulations using CT-INT found a critical point at $V_{c}/t=1.356(1)$ with critical exponents $\eta= 0.302(7)$ and $\nu= 0.80(3)$. \cite{Wang:2014ib}  Although CT-INT is free from the time-discretization error, its cubic scaling with inverse temperature $\beta$ limited these simulations to inverse temperatures $\beta t \leq 20$. To access the quantum critical point from a finite temperature simulation $\beta$ was scaled linearly with the linear extent of the system, assuming a dynamical critical exponent $z=1$. In Sec. 
\ref{sec:tvresults}  we will, as a first application of the projector LCT-INT algorithm, use it to directly address the quantum critical point of model (\ref{eq:Ham}) at zero-temperature and check our previous findings.

\section{Algorithm \label{sec:algorithm}}

\subsection{General Description}
In a projector QMC calculation one obtains the ground state wave function using 
imaginary time projection of a trial wave function $ |\Psi_{T} \rangle$ and calculate ground state observables as~\cite{Sugiyama:1986vt, Sorella:1989uk}

\begin{equation}
\langle {\hat{O}} \rangle = \frac{\langle\Psi_{T} |  e^{-\Theta \hat{H}/2} \,\hat{O} \, e^{-\Theta \hat{H}/2} |\Psi_{T} \rangle }{\langle\Psi_{T} | e^{-\Theta \hat{H}}|\Psi_{T} \rangle}. 
\label{eq:projection}
\end{equation}
For any $|\Psi_{T}\rangle$ with non-vanishing overlap with the true ground state, Eq.~(\ref{eq:projection}) approaches the ground state expectation in large $\Theta$ limit. In this paper, we choose $|\Psi_{T}\rangle$ as a single Slater determinant, 
\begin{equation}
|\Psi_{T}\rangle = \prod_{j=1}^{N_{P}} \left(\sum_{i=1}^{N} P_ {ij} \hat{c}^{\dagger}_{i}\right) |0\rangle,  
\end{equation}
where $N$ is number of sites, $N_{P}$ is the number of particles, and $P$ is a $N\times N_{P}$ rectangular matrix.\footnote{We will consider the half-filled case with $N_{P} = N/2$ in the following for a sign-problem free simulation.} 

%
%

Instead of breaking the projection operator into small time steps as done in discrete time algorithms,~\cite{Blankenbecler:1981vj, Sugiyama:1986vt, Sorella:1989uk, White:1989wh, Loh:1992, Assaad:2008hx} the continuous-time QMC formalism writes the projection operator in an interaction representation


\begin{equation}
e^{- \Theta \hat{H}} =  e^{- \Theta \hat{H}_{0}}\, \mathcal{T} \exp \left[-\int_{0}^{\Theta}\! e^{\tau \hat{H}_{0}}\hat{H}_{1}e^{-\tau \hat{H}_{0}} \,\mathrm{d}\tau \right]. 
\label{eq:interaction picture}
\end{equation}

After a Taylor expansion of the exponential and time ordering the terms~\footnote{This rearrangement cancels the factorial factor $1/k!$.} the denominator of Eq.~(\ref{eq:projection}) reads
\begin{widetext}
\begin{eqnarray}
\langle\Psi_{T} | e^{-\Theta \hat{H}}|\Psi_{T} \rangle 
  & = & \sum_{k=0}^{\infty} {(- 1)^k} \int_0^{\Theta} \mathrm{d}  \tau_1
  \int_{\tau_1}^{\Theta} \mathrm{d} \tau_2 \ldots \int_{\tau_{k - 1}}^{\Theta}\mathrm{d} \tau_k\,
  \langle \Psi_T | e^{- (\Theta - \tau_k) \hat{H}_{0}} \hat{H}_{1}  \ldots \hat{H}_{1}e^{- (\tau_2 - \tau_{1}) \hat{H}_{0}}\hat{H}_{1}e^{- \tau_1 \hat{H}_{0}} | \Psi_T \rangle. 
  \label{eq:expansion}
\end{eqnarray}
\end{widetext}

In the CT-INT and the CT-AUX methods,~\cite{Rubtsov:2005iw, Assaad:2007be, Gull:2008cm} one applies Wick's theorem to the integrand of Eq.~(\ref{eq:expansion}) and expresses it as a determinant of a matrix whose size is proportional to the expansion order $k$. The subsequent simulation modifies the matrix with $\mathcal{O}(k^{2})$ operations per Monte Carlo step. Since it takes $k$  Monte Carlo steps to generate an uncorrelated sample, these methods~\cite{Rubtsov:2005iw, Assaad:2007be, Gull:2008cm} scale \emph{cubically} with the average expansion order $\langle k\rangle$. As $\langle k\rangle$ increases linearly with inverse temperature $\beta$ (or $\Theta$ in the ground state projection scheme) and interaction strength,~\cite{Rubtsov:2005iw, Assaad:2007be, Gull:2008cm} this unfavorable cubic scaling limits the applicability of these methods for lattice models at low temperature and strong interactions. Here we instead use the LCT-INT algorithm~\cite{Mauro} to achieve \emph{linear scaling} with respect to the average expansion order. The algorithm scales as $\Theta V N^{3}$, similar to the  BSS algorithm.~\cite{Blankenbecler:1981vj, Loh:1992, Assaad:2008hx}

To proceed, we first express the interaction term $\hat{H}_{1}$ through exponentials of bilinear fermion operators. Traditionally this is done via Hubbard-Stratonovich transformation, at the cost of introduction of auxiliary fields.~\cite{PhysRevLett.82.4155, Gull:2008cm} Here we adopt a simpler approach based on the operator identity $\hat{n}_{i} = \frac{1}{2} (1- e^{i\pi \hat{n}_{i}})$. The interaction term Eq.~(\ref{eq:V}) can then be expressed as 

\begin{equation}
\hat{H}_{1} =\frac{V}{4}\sum_{\langle i,j\rangle}  e^{i\pi(\hat{n}_{i}+\hat{n}_{j})}. 
\label{eq:Vasexp}
\end{equation}
This reformulation works for any density-density interaction\footnote{It can be understood as a special case of the auxiliary field decomposition, where one sets $\alpha=\beta=-2$ in Eq.~(4) of Ref. \onlinecite{Rombouts:1998tf}, or chooses the shift parameter $\mu=\frac{-U\beta N_{s}}{4}$ in Eq.~(6) of Ref.  \onlinecite{PhysRevLett.82.4155}, or uses $\alpha =0$ in Eq.~(10) of Ref. \onlinecite{PhysRevE.79.057701}. This special choice of parameter breaks the connection between the CT-INT and the discrete time BSS algorithm~\cite{PhysRevE.79.057701} and explains why the occurrence of a sign problem can be different in these algorithms.} and is crucial to respect the symmetry of the model (\ref{eq:Ham}),  ensuring a sign problem free QMC simulation. Substituting Eq.~(\ref{eq:Vasexp}) into Eq.~(\ref{eq:expansion}), the integrand is recognized as a sum of determinants
\begin{widetext}
\begin{eqnarray}
\langle\Psi_{T} | e^{-\Theta \hat{H}}|\Psi_{T} \rangle
 & = &  \sum_{k=0}^{\infty}\left(\frac{- V}{4}\right)^{k} \sum_{\langle {i}_{1},{j}_{1}\rangle} \sum_{\langle {i}_{2},{j}_{2}\rangle} \ldots \sum_{\langle {i}_{k},{j}_{k}\rangle} \int_0^{\Theta} \mathrm{d}  \tau_1
  \int_{\tau_1}^{\Theta} \mathrm{d}  \tau_2 \ldots \int_{\tau_{k - 1}}^{\Theta} \mathrm{d}  \tau_k \times \nonumber \\
 && \det\left[P^{\dagger}e^{-(\Theta-\tau_{k}) K} {X}(i_{k},j_{k})  \ldots  {X}( {i}_{2}, {j}_{2}) e^{-(\tau_{2}-\tau_{1}) K} {X}({i}_{1},{j}_{1}) e^{-\tau_{1}K} P\right],  
 \label{eq:det}
\end{eqnarray}
\end{widetext}
where $K$ is defined in Eq.~(\ref{eq:K}). The vertex matrix $X(i,j)$ is an $N\times N$ diagonal matrix defined for nearest neighbors $i$ and $j$ whose non-zero elements are 
\begin{equation}
X(i,j)_{ll} =\begin{cases}
-1, & \text{$l=i$ or $l=j$}, \\ 
1, & \text{otherwise}.
  \end{cases}
 \label{eq:vertexmatrix} 
\end{equation}
Its form follows immediately from  Eq.~(\ref{eq:Vasexp}). 
In the following we denote the matrix product from imaginary time $\tau$ to $\tau^{\prime}>\tau$ as a propagator 
\begin{equation}
{B}(\tau^{\prime}, \tau) = e^{-(\tau^{\prime}-\tau_{m}) K} X(i_{m},j_{m})  \ldots  X({i}_{l},{j}_{l}) e^{-(\tau_{l}-\tau) K}, 
\label{eq:Bpropagator}
\end{equation}
where $\tau_{l}$ and $\tau_{m}$ are the imaginary time locations of the first and the last vertices in the time interval. The corresponding vertex matrices are $X({i}_{l},{j}_{l}) $ and $X(i_{m},j_{m})$ respectively. If there is no vertex in the time interval, the propagator simply reads ${B}(\tau^{\prime}, \tau) =  e^{-(\tau^{\prime}-\tau) K}$.

\begin{figure}[b]
\centering 
\includegraphics[width=8cm]{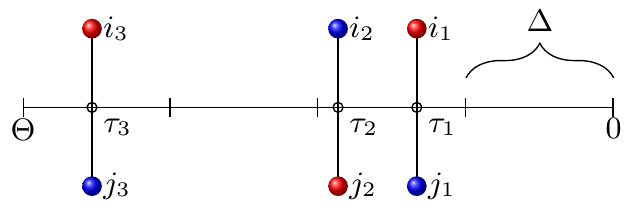}
\caption{A configuration with $k=3$ vertices. We divide the imaginary time axis into intervals of size $\Delta$ and sweep through them sequentially. In each interval we propose updates that either insert or to remove a vertex. Measurement is performed at the center of the imaginary time $\tau = \Theta/2$.}
\label{fig:configuration}
\end{figure}

We then expand the nominator of Eq.~(\ref{eq:projection}) similarly and write the expectation value in a form suitable for Monte Carlo sampling, 
\begin{equation}
\langle \hat{O} \rangle = \frac{\sum_{\mathcal{C}} w(\mathcal{C}) \langle \hat{O}\rangle_{\mathcal{C}, \Theta/2} }{\sum_{\mathcal{C}} w(\mathcal{C}) }  = \left\langle \langle \hat{O}\rangle_{\mathcal{C}, \Theta/2}  \right\rangle_\mathrm{MC},
\label{eq:MC}
\end{equation}
where the configuration $\mathcal{C}$ denotes a point in the summation and integration domain of Eq.~(\ref{eq:det}) and $\langle \ldots \rangle_\mathrm{MC}$  denotes Monte Carlo averaging according to the configuration weights  $w(\mathcal{C})$. 

For $k$ vertices a configuration
\begin{equation}
\mathcal{C} = \{(\tau_{1}; i_{1}, j_{1}), (\tau_{2}; i_{2}, j_{2}) ,  \ldots,  (\tau_{k}; i_{k}, j_{k})\}. 
\end{equation}
consists of ordered times $0\le\tau_{1}<\tau_{2}<\ldots < \tau_{k}<\Theta$ and corresponding pairs of nearest neighbor sites $( i_{1},j_{1} )$, $( i_{2},j_{2} )$ \ldots $ ( i_{k}, j_{k})$. 
An example of a configuration with three vertices is shown in  Fig.~\ref{fig:configuration}. 

Using Eq.~(\ref{eq:Bpropagator}) the weight of a configuration is expressed as   
\begin{equation}
w(\mathcal{C}) = \left(\frac{- V}{4}\right)^{k}\det[P^{\dagger}{B}(\Theta, 0)P]\, \mathrm{d}\tau_{1} \ldots \mathrm{d}\tau_{k} .   
\label{eq:weight}
\end{equation}
Since this weight is, up to a constant factor, identical to the weight of such a configuration in a CT-INT calculation~\cite{Huffman:2014fj, Wang:2014ib} at zero temperature,~\cite{Assaad:2007be} these methods have identical sign problems.

The quantum mechanical average $\langle \hat{O}\rangle_{\mathcal{C}, \tau}$ of an operator $\hat{O}$ inserted into a configuration $\mathcal{C}$ at imaginary time $\tau$:
\begin{align}
&\langle \hat{O} \rangle_{\mathcal{C}, \tau} =  \nonumber \\
&\frac{\langle \Psi_T | e^{- (\Theta - \tau_k) \hat{H}_{0}}e^{i\pi(\hat{n}_{i_{k}}+\hat{n}_{j_{k}})} \ldots \hat{O} \ldots  e^{i\pi(\hat{n}_{i_{1}}+\hat{n}_{j_{1}})}  e^{- \tau_1 \hat{H}_{0}} | \Psi_T \rangle  }{ \langle \Psi_T | e^{- (\Theta - \tau_k) \hat{H}_{0}}e^{i\pi(\hat{n}_{i_{k}}+\hat{n}_{j_{k}})}   \ldots e^{i\pi(\hat{n}_{i_{1}}+\hat{n}_{j_{1}})}  e^{- \tau_1 \hat{H}_{0}} | \Psi_T \rangle }\nonumber 
\end{align}
can be evaluated using Wick's theorem since $\hat{O}$ is sandwiched between two Slater determinants. 

\subsection{Monte Carlo Sampling}
In this section we first explain how to sample using the weights Eq.~(\ref{eq:weight}) and then discuss efficient ways to perform update and measurement using equal-time Green's function.

\subsubsection{General Procedure}
To sample configurations $\mathcal{C}$ according to the weight $w(\mathcal{C})$ we use the Metropolis-Hastings algorithm.~\cite{Metropolis:1953in,Hastings:1970aa} Starting from a  configuration $\mathcal{C}$ we propose to move to a new configuration $\mathcal{C}^{\prime}$ with an a-priori probability $A(\mathcal{C}\rightarrow\mathcal{C}^{\prime})$. The new configuration is then accepted with probability $
p(\mathcal{C}\rightarrow\mathcal{C}^{\prime}) = \min\left\{1, r(\mathcal{C}\rightarrow\mathcal{C}^{\prime})\right\},
$ where the acceptance ratio $r$ is 
\begin{equation}
 r(\mathcal{C}\rightarrow\mathcal{C}^{\prime}) = \frac{w(\mathcal{C}^{\prime})A(\mathcal{C}^{\prime}\rightarrow\mathcal{C}) }{w(\mathcal{C})A(\mathcal{C}\rightarrow\mathcal{C}^{\prime})}. 
\end{equation}

To facilitate fast computation of the acceptance rate, we divide the imaginary time axis into intervals of size $\Delta$, as shown in Fig.~\ref{fig:configuration}. We focus our updates on one interval at a time, proposing several times to either insert or to remove an existing vertex at time $\tau$ for site indices $\langle i,j\rangle$.  
Sweeping through the intervals we achieve ergodicity. While such sequential updates violate the detailed balance condition, a global balance condition is still restored as long as the updates within each interval satisfy \emph{local} detailed balance.~\cite{Manousiouthakis:1999gc}

Using short hand notations,~\cite{Loh:1992, Assaad:2008hx}
\begin{equation}
L(\tau)= P^{\dagger} B(\Theta, \tau) \quad \mathrm{and}\quad  R (\tau)= B(\tau, 0) P, 
\label{eq:LR}
\end{equation}
the insertion or removal changes $R$ to $R^{\pm} = X(i,j)^{\pm 1} R$. Note that for the model Eq.~(\ref{eq:Ham}) studied here $R^+=R^-$ because $X(i,j)^{-1} = X(i,j)$, but the general Monte Carlo scheme does not rely on this property. The acceptance ratios are

\begin{eqnarray}
  r_{\mathrm{add}} & = & - \frac{\det(LR^{+})}{\det(LR)}  \times\frac{ VN_{b}\Delta}{4(n+1)} ,  \label{eq:add}
  \\
  r_{\mathrm{remove}} & = &- \frac{\det(LR^{-})}{\det(LR)}\times  \frac{4  n} {N_{b}V \Delta}  , 
  \label{eq:remove}
\end{eqnarray}
where $N_{b}$ is the number of interacting bonds of the system, $\Delta$ is the length of the time interval  on which we propose updates and $n$ is the number of vertices in this interval. While insertion and removal updates are sufficient to ensure ergodicity of the sampling, one can nevertheless implement additional updates to improve the sampling efficiency, such as change the site index of a vertex (see  Appendix \ref{sec:siteshift}). 

 After a full sweep through all the intervals, we measure the expection values of observables close to the center $\tau = \Theta /2$. 

\subsubsection{Fast Update Using Equal-time Green's Function}

Crucial for the performance of the algorithm is a fast calculation of the acceptance ratios Eqs.~(\ref{eq:add}-\ref{eq:remove}). They can be efficiently computed from the equal time Green's functions ${G}_{lm}(\tau) = \langle \hat{c}_{l} \hat{c}_{m}^{\dagger} \rangle_{\mathcal{C}, \tau}$, which  in matrix form reads~\cite{Loh:1992, Assaad:2008hx}
\begin{equation}
{G}(\tau)= \mathbb{I}- R(\tau) [L(\tau)R(\tau)]^{-1} L(\tau). 
\label{eq:G}
\end{equation}
The determinant ratio in Eqs.~(\ref{eq:add}-\ref{eq:remove}) can be expressed using the Green's function as~\cite{Loh:1992, Assaad:2008hx}
\begin{widetext}
\begin{eqnarray}
- \frac{\det(LR^{\pm})}{\det(LR)}  & = & -\det\left\{\mathbb{I} + [{X}(i, j)^{\pm 1} -\mathbb{I} ] (\mathbb{I} -{G})\right\}= -\det\left(\begin{array}{cc}1-2(1-G_{ii}) & 2G_{ij} \\ 2G_{ji}& 1-2(1-G_{jj})\end{array}\right) \nonumber \\
 &=& 4 {G}_{ij} {G}_{ji}
 \label{eq:detratio}
\end{eqnarray}
\end{widetext}

The second equality follows from  $[{X}({i},{j})^{\pm 1}-\mathbb{I}]_{lm} = - 2\delta_{li}\delta_{im} -2\delta_{lj} \delta_{jm}$. With appropriately chosen trial wave function, the equal-time Green's function of our model  (\ref{eq:Ham}) has an important symmetry property which we prove in Appendix~\ref{sec:symmetry}:
\begin{equation}
{G}_{ji}(\tau) = \delta_{ij} - \eta_{i}\eta_{j} {G}_{ij}(\tau), 
\label{eq:symmetry}
\end{equation} 
where $\eta_{i}= \pm 1$ for site $i$ belongs to the A(B) sublattice. Similar to the case of CT-INT,~\cite{Huffman:2014fj, Wang:2014ib} Eq.~(\ref{eq:symmetry}) implies $G_{ii}=\frac{1}{2}$ and ${G}_{ij} = {G}_{ji}$ for nearest neighbors. With this we can further simplify the determinant ratios to $4{G}_{ij} {G}_{ji} = 4{G}_{ij}^2$. Since the remaining factors in Eqs. (\ref{eq:add}-\ref{eq:remove}) are all positive, there is no sign problem in the simulation of the model (\ref{eq:Ham}).~\cite{Huffman:2014fj, Wang:2014ib} 

If a proposed Monte Carlo move is accepted, we update the Green's function to $G^{\pm} = \mathbb{I}- R^{\pm} (LR^{\pm})^{-1} L$ using  
 \begin{equation}
 {G}^{\pm}_{lm} =  {G}_{lm} - \frac{ {G}_{l j} ( {G}_{i m} - \delta_{im})}{ {G}_{ij}} - \frac{ {G}_{l i} ( {G}_{j m} - \delta_{jm})}{ {G}_{ji}}.
\label{eq:updateG}
\end{equation}
For a proof see Appendix~\ref{sec:updateG}.

In the Monte Carlo updates we always keep track of the Green's function at the imaginary time $\tau$ for which we propose an update. For the next Monte Carlo move we need to propagate the Green's function to a different imaginary time $\tau^{\prime}$, which can be done by the following similarity transformation (assuming $\tau^{\prime}>\tau$) 
\begin{equation}
 {G}(\tau^{\prime}) = {B}(\tau^{\prime} , \tau) {G}(\tau) \left [{B}(\tau^{\prime} , \tau) \right ]^{-1}. 
 \label{eq:wrap}
\end{equation}
Propagating the Green's function using the tricks discussed in Sec.~\ref{sec:optimization}, Eq.~(\ref{eq:wrap}) is more efficient than calculating $G(\tau^{\prime})$ from scratch using Eq.~(\ref{eq:G}). 

%
%

\subsubsection{Observables}

All expectation values $\langle \hat{O}\rangle_{\mathcal{C}, \tau}$ can be related to the Green's function using  Wick's theorem. For example, the density-density correlation function can be expressed as $\langle \hat{n}_{l} \hat{n}_{m}\rangle_{\mathcal{C}, \tau} = (1- {G}_{ll})(1- {G}_{mm}) + (\delta_{lm}-  {G}_{ml})  {G}_{lm}$. Using Eq.~(\ref{eq:symmetry}), the density-density correlation functions is measured as
\begin{eqnarray}
C(l,m) & = & \left\langle \left(\hat{n}_{l}-\frac{1}{2}\right)  \left(\hat{n}_{m}-\frac{1}{2}\right) \right\rangle \nonumber \\
&= & \eta_{l} \eta_{m}  \left \langle {G_{lm}^{2}\left(\frac{\Theta}{2}\right) } \right\rangle_\mathrm{MC},
\label{eq:Cr}
\end{eqnarray}
the staggered CDW structure factor as
\begin{eqnarray}
M_{2} & = & \frac{1}{N^{2}}\sum_{l,m } \eta_{l} \eta_{m} \left\langle \left(\hat{n}_{l}-\frac{1}{2}\right) \left(\hat{n}_{m}-\frac{1}{2}\right) \right\rangle \nonumber \\
& = & \frac{1}{N^{2}}  \sum_{l,m} \left \langle {G_{lm}^{2}\left(\frac{\Theta}{2}\right) } \right\rangle_\mathrm{MC}, \label{eq:M2} 
\end{eqnarray}
and the kinetic energy and interaction energy as  
\begin{eqnarray}
\langle \hat{H}_{0}\rangle &  =& -\Tr\left( K\left\langle G\left(\frac{\Theta}{2}\right) \right\rangle_\mathrm{MC}\right), \\
\langle \hat{H}_{1}\rangle & = & - V \sum_{\langle l,m \rangle} \left \langle {G_{lm}^{2}\left(\frac{\Theta}{2}\right) } \right\rangle_\mathrm{MC} . 
\end{eqnarray}


Another useful observable is the average expansion order 
\begin{equation}
\langle k \rangle = - \left \langle \int_{0}^{\Theta} \langle \hat{H}_{1}\rangle_{\mathcal{C}, \tau} \mathrm{d}\tau \right \rangle_\mathrm{MC}. 
\label{eq:averagek}
\end{equation}
Since there is no translational invariance along the imaginary time axis, $\langle k \rangle$ is not directly related to the interaction energy $\langle \hat{H}_{1}\rangle$ as it is in the finite temperature case.~\cite{Rubtsov:2005iw} Nevertheless, Eq.~(\ref{eq:averagek}) still suggests $\langle k \rangle\sim \Theta V N$, i.e. the average number of vertices scales linearly with the projection time, the interaction strength and the system size. The fact that we are dealing with $k$ of $N\times N $ matrices compared to the single $2k\times 2k$ matrix of the CT-INT case~\cite{Wang:2014ib} allows LCT-INT to achieve an $\mathcal{O}(\Theta V N^{3})$  scaling, as we will discuss in the next section. 

\subsection{Algorithm Optimization and Complexity~\label{sec:optimization}}
Achieving the same scaling of  $\mathcal{O}(\Theta V N^{3})$ as in the BSS algorithm requires a careful implementation, for which an optimal choices of single particle basis and splitting imaginary time into intervals is crucial.  

\subsubsection{Optimal Single-Particle Basis \label{sec:eigenbasis}}

The main computational effort in performing the  Monte Carlo updates is the propagation of the Green's function to a new imaginary-time using Eq.~(\ref{eq:wrap}). Implemented na\"ively this involves dense matrix-matrix multiplication and requires $\mathcal{O}(N^{3})$ operations, while the cost of the calculation of the determinant ratio in Eq.~(\ref{eq:detratio}) is $\mathcal{O}(1)$ and update of the  Green's function Eq.~(\ref{eq:updateG}) is $\mathcal{O}(N^{2})$. 

This unfavorable scaling can be circumvented by working in the eigenbasis of the noninteracting Hamiltonian.~\cite{Mauro} In this way all the computations for MC steps Eqs.~(\ref{eq:wrap},\ref{eq:detratio},\ref{eq:updateG}) can be performed with complexity $\mathcal{O}(N^{2})$. 

For this we have to use basis-transformed Green's functions $\tilde{G} = U^{\dagger} G U$, where $U$ are the eigenvectors of the single-particle Hamiltonian ${U}^{\dagger} {K} {U} = \diag(E_{1}, E_{2}, \ldots, E_{N})$. The basis change modifies the propagators to
\begin{eqnarray}
\left(U^{\dagger}e^{-\tau K}U\right)_{lm} & = &e^{-E_{l}\tau}\delta_{lm} \label{eq:eigenbasisK} , \\
\left(U^{\dagger}X (i,j) U\right)_{lm}& =& \delta_{lm} -2 {U}^{\dagger}_{li} {U}_{im} -2 {U}^{\dagger}_{lj} {U}_{jm}.
\label{eq:eigenbasisV}
\end{eqnarray}
In this basis the multiplication of $\tilde{G}$ by either Eq.~(\ref{eq:eigenbasisK}) or Eq.~(\ref{eq:eigenbasisV}) requires only $\mathcal{O}(N^{2})$ operations instead of $\mathcal{O}(N^{3})$.~\cite{Mauro} The disadvantage is that now the calculation the determinant ratio Eq.~({\ref{eq:detratio}}) is slightly more expensive.    
However since we only need one matrix element $G_{ij} = (U\tilde{G}U^{\dagger})_{ij}$ which can be calculated using matrix-vector multiplication and vector inner-products, this $\mathcal{O}(N^{2})$ overhead will not affect the overall scaling of the algorithm. Similarly, updating of the Green's function in the eigenbasis also keeps an $\mathcal{O}(N^{2})$ scaling (see Appendix~\ref{sec:eigen}). 

Working in the  eigenbasis of the noninteracting Hamiltonian does not increase the complexity of the measurements either. One can choose to  measure single particle observables in the eigenbasis and perform a basis rotation afterwords. Alternatively one can rotate $\tilde{G}$ back to $G$ with $\mathcal{O}(N^{3})$ operations for each measurement. Since measurements are performed only after a full sweep through all intervals, this does not affect the overall scaling of the algorithm. For many physical observables of interest we only need ${G}_{ij}$ for neighboring sites $\langle i,j\rangle$ or for fixed site $i$ because of translational invariance, which would reduce the required basis transformation to just $\mathcal{O}(N^{2})$ operations. 

\subsubsection{Optimal Interval Size}

Finally we show that by choosing the number of intervals $M=\Theta/\Delta$ proportional to the average number of vertices one can achieve an overall $\mathcal{O}(\Theta V N^{3})$ scaling in the algorithm. For each MC update, we need to propagate the Green's function from some time $\tau$ to another time $\tau^{\prime}$ in the same interval. This will on average pass through $\frac{|\tau-\tau^{\prime}|}{\Theta}\langle k \rangle$ existing vertices, which is of order $\mathcal{O}(\langle k \rangle/M)$. As we need $\mathcal{O}(N^{2})$ operations to pass through each vertex and $\mathcal{O}(N^{2})$ for calculating the acceptance rate and for the actual update, one Monte Carlo step requires $\mathcal{O}(\max\{1, \langle k \rangle/M\} N^{2})$ operations, where the max function accounts for the case of an empty interval. 

A sweep through all intervals and updating $\langle k \rangle $ vertices results in an overall number of $\mathcal{O} (\max\{1, { \langle k\rangle}/M\}N^{2}\langle k \rangle) $ operations. By choosing the number of intervals $M\sim \langle k\rangle$ we can achieve an optimal scaling $\mathcal{O} (\langle k\rangle N^{2}) = \mathcal{O} (\Theta V N^{3}) $. This should be compared to the scaling of other continuous-time methods which scale as $\langle k\rangle^{3} \sim \Theta^{3} V^{3} N^{3}$.~\cite{Rubtsov:2005iw, Assaad:2007be, Gull:2008cm, Gull:2011jd}

\subsection{Numerical Stabilization}
As in the BSS algorithm the multiplication of the Green's function with the propagator $B(\tau^{\prime},\tau)$ for large imaginary-time suffers from numerical instabilities because the matrix multiplication mixes large and small scales in the propagator. We stabilize the calculation following a similar approach as used for the BSS method.~\cite{White:1989wh} The following discussion largely follows Refs.~\onlinecite{Loh:1992, Assaad:2008hx} with the difference that our stabilization is done in continuous time and in the eigenbasis of the single-particle Hamiltonian.  


To avoid accumulation of numerical errors we need to regularly recompute the Green's function using $\tilde{G} = \mathbb{I}- U^{\dagger}R (LR)^{-1} L U$, which requires us to have fast access to the matrices $U^{\dagger}R$ and $LU$ in a numerical stable way. We thus divide the imaginary time axis into $I$ intervals where within each interval the propagation is well-conditioned. The interval length is set by the inverse bandwidth and is independent of the total projection time $\Theta$. These intervals  are different from the (shorter) intervals used for MC updates discussed above. 

At the interval boundaries (corresponding to imaginary times $\tau^{\ell=0}=0,\ldots,\tau^{\ell=I}=\Theta$) we
 store $I+1$ matrices $S^{\ell}$. Depending on the current imaginary time $\tau$ of the Monte Carlo sweep, they hold the matrix product either to the right or to the left, 
\begin{equation}
S^{\ell} =\begin{cases}
U^{\dagger} R(\tau^{\ell}), & \text{if $\tau> \tau^{\ell}$}, \\ \\
L(\tau^{\ell})U, & \text{otherwise}.
  \end{cases}
\end{equation}
On the rightmost and leftmost boundaries $S^{\ell=0}=U^{\dagger}P$ and $S^{\ell=I}=P^{\dagger}U$. The matrix $S^{\ell}$ is updated whenever we cross the interval boundary $\tau^{\ell}$ in the sweep along the imaginary time axis.
During a sweep from $\tau=0$ to $\Theta$, we multiply the  propagator $\tilde{B}(\tau^{\ell},\tau^{\ell-1})\equiv U^{\dagger}B(\tau^{\ell},\tau^{\ell-1}) U$ with $S^{\ell-1}$ to update $S^{\ell}=\tilde{B}(\tau^{\ell},\tau^{\ell-1})S^{\ell-1}$. In the backward sweep from $\tau=\Theta$ to $0$, we update $S^{\ell}$ to  $S^{\ell}= S^{\ell+1}\tilde{B}(\tau^{\ell+1},\tau^{\ell})$. 

We still need to stabilize the calculation of $S^{\ell=0,\ldots,I}$ themselves. Performing a singular-value-decomposition (SVD) \begin{eqnarray}
U^{\dagger}R & = & U_{R} D_{R} V_{R}, \label{eq:UdaggerR}\\ 
LU & = & U_{L} D_{L} V_{L}, \label{eq:LU} 
\end{eqnarray}
the different scales only appear in the diagonal matrices of singular values $D_L$ and $D_R$. Since $\tilde{G} = \mathbb{I}- U_{R} (V_{L} U_{R}) ^{-1} V_{L}$ only depends on  the well-conditioned matrices $U_{R}$ and $V_{L}$, it is sufficient to keep track of them instead of the full matrix products. Therefore before updating $S^{\ell}$ we can perform an SVD on the matrix product $\tilde{B}(\tau^{\ell},\tau^{\ell-1})S^{\ell-1}$ or $S^{\ell+1}\tilde{B}(\tau^{\ell+1},\tau^{\ell})$ and only store $U_{R}$ or $V_{L}$. 
 
Using these stored matrices $S^{\ell=0,\ldots,I}$ we can easily  recompute the Green's function at any imaginary time. To compute $\tilde{G}(\tau)$ for  $\tau^{\ell+1}>\tau >\tau^{\ell}$ we can use the matrices $S^{\ell+1}$ and $S^{\ell}$ and calculate $U_{R}= \tilde{B}( \tau, \tau^{\ell})S^{\ell}$ and $V_{L}=S^{\ell+1} \tilde{B}( \tau^{\ell+1}, \tau) $. The Green's function is then  recomputed by $\tilde{G} = \mathbb{I}- U_{R} (V_{L}U_{R}) ^{-1} V_{L}$. 



In the simulation we monitor the difference of the stabilized $\tilde{G}$ and the old one to dynamically adjust the frequencies of the SVD and the recomputation of $\tilde{G}$. It turns out both frequencies are mainly set by the inverse bandwidth and are independent of the system size or the total projection time. Typically we need to perform one of such stabilizations for a propagation time $\tau \sim 1/t$. Since each of these stabilization steps costs $\mathcal{O}(N^{3})$ due to the SVD or matrix inverse and we need to perform $\mathcal{O}(\Theta)$ of them per sweep, it ends up with a scaling of $\mathcal{O}(\Theta N^{3})$, conforming with the overall scaling of the algorithm.


%

\subsection{Calculation of the Renyi Entanglement Entropy}
Quantum information based measures  play an increasing role in the identification of quantum phases and phase transitions.\cite{PhysRevB.81.064439, Isakov:2011fz, Jiang:2012dw,PhysRevLett.110.260403} In particular, Refs.~\onlinecite{Grover:2013cs, Broecker:2014fl, Wang:2014vn} devised measurements of the Renyi entanglement entropy in determinantal QMC simulations. 

Since in the present algorithm the many-body ground state wave function is expressed as a sum of Slater determinants, the derivations of Ref.~\onlinecite{Grover:2013cs} concerning the reduced density matrix hold. In particular, the rank-2 Renyi entanglement entropy $S_{2}= -\ln\Tr (\hat{\rho}_{A}^{2})$ of a region $A$ can be calculated using, 

\begin{equation}
e^{-S_{2}} = \frac{\sum_{\mathcal{C,C}^{\prime}}w(\mathcal{C}) w(\mathcal{C}^{\prime}) \det\left[ G_{A} G_{A}^{\prime} + (\mathbb{I}-G_{A})(\mathbb{I}-G_{A}^{\prime})\right] }{\sum_{\mathcal{C},\mathcal{C}^{\prime}} w(\mathcal{C}) w(\mathcal{C}^{\prime}) },
\label{eq:S2}
\end{equation} 
where $\mathcal{C}$ and $\mathcal{C}^{\prime}$ are configurations of two independent replicas and $G_{A}$, $G_{A}^{\prime}$ are the corresponding Green's function restricted to the region $A$. 

Although the estimator Eq.~(\ref{eq:S2}) is easy to implement, we observe it shows large fluctuations at strong interaction.~\cite{2014PhRvB..89l5121A, Anonymous:XiXakTeu, Broecker:2014fl} We leave a discussion of extended ensemble simulations~\cite{Humeniuk:2012cqa, Broecker:2014fl, Wang:2014vn} in the LCT-INT formalism for a future study.

\subsection{Direct Sampling of Derivatives  \label{sec:derv}}

An advantage of continuous-time algorithms over discrete-time algorithms is that the Monte Carlo weight Eq.~(\ref{eq:weight}) are homogeneous functions of the interaction strength $V$. This allows to directly sample the derivatives of any observable with respect to~$V$ using its covariance with the expansion order $k$, 
\begin{equation}
\frac{\partial{\langle \hat{O}\rangle}}{\partial V} = 
\left\langle \frac{\partial{ \hat{O}}}{\partial V}\right\rangle+ \frac{1}{V} \left(\langle \hat{O} k \rangle -  \langle \hat{O}\rangle \langle  k \rangle \right). 
\label{eq:derv}
\end{equation}
Higher order derivatives can be sampled in a similar way. Derivatives are useful for discovering quantum phase transitions and locating critical points.

 In discrete time approaches calculation of such observables either relies on the Hellmann-Feynman theorem,~\cite{Hellmann1937, PhysRev.56.340} which is limited to the first order derivative of the total energy $ \langle\hat{H}\rangle$,~\cite{Hohenadler:2012is,Hohenadler:2014bz} or requires noisy numerical differentiation of Monte Carlo data.~\cite{Meng:2010gc}

\begin{figure}[t]
\includegraphics[width=9cm]{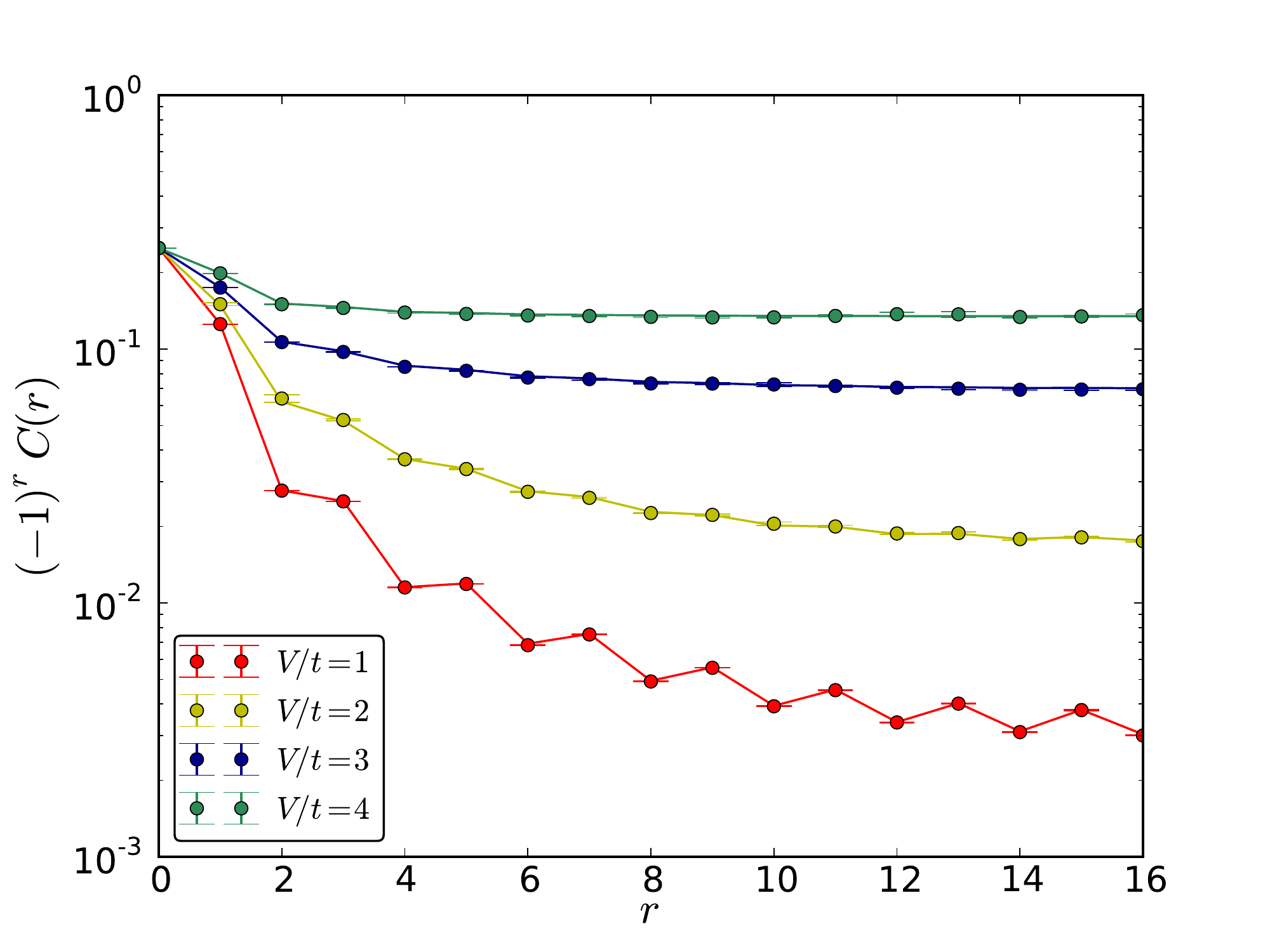}
\caption{Density-density correlation function of a $32$-site chain with periodic boundary condition. Solid lines are DMRG results. }
\label{fig:nncorr}
\end{figure}

\section{Results \label{sec:results}}

We finally present results obtained with our algorithm, starting with benchmarks that demonstrate the correctness before presenting new results regarding the quantum critical point of the model Eq.~(\ref{eq:Ham}).

\subsection{Benchmarks}

For all of our results we use a projection time $\Theta t=40$ and use ground state of the noninteracting Hamiltonian $\hat{H}_{0}$ as the trial wave function. In case of degenerate noninteracting ground states, we take as  trial wave function the ground state of a system with anti-periodic boundary condition in $x$-direction and periodic boundary condition along the $y$-direction.

We start by showing results for a periodic chain. Figure~\ref{fig:nncorr} shows the density-density correlation function of a periodic chain compared with results from density matrix renormalization group (DMRG) calculations, where $C(r)$ is averaged over all pairs in Eq.~(\ref{eq:Cr}) with the same distances $r$. At moderate computational cost we can perfectly reproduce the exact ground state quantities using projection LCT-INT (filled dots). 

\begin{figure}[t]
\includegraphics[width=9cm]{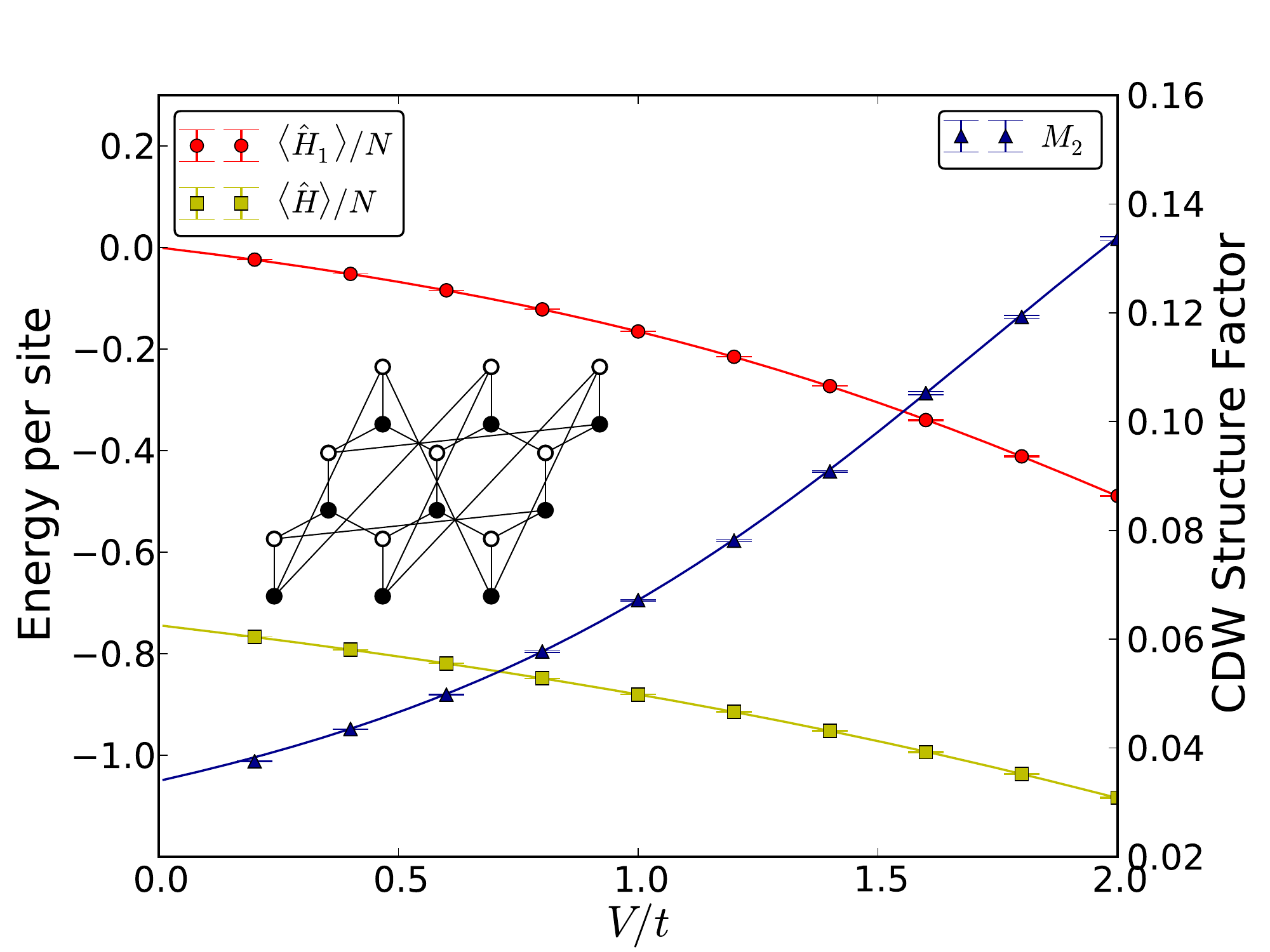}
\caption{Interaction energy (red dots), ground state energy (yellow squares) and CDW structure factor $M_{2}$ (blue triangles) of an $N=18$ site honeycomb lattice shown in the inset. Solid lines are exact diagonalization results.}
\label{fig:benchmark}
\end{figure}

Figure~\ref{fig:benchmark}  compares the results obtained with our algorithm to exact diagonalization (solid lines) for an $N=18$ site honeycomb lattice. Our method correctly produces ground state results for the total energy, interaction energy as well as the staggered density structure factor. 

\begin{figure}[t]
\includegraphics[width=9cm]{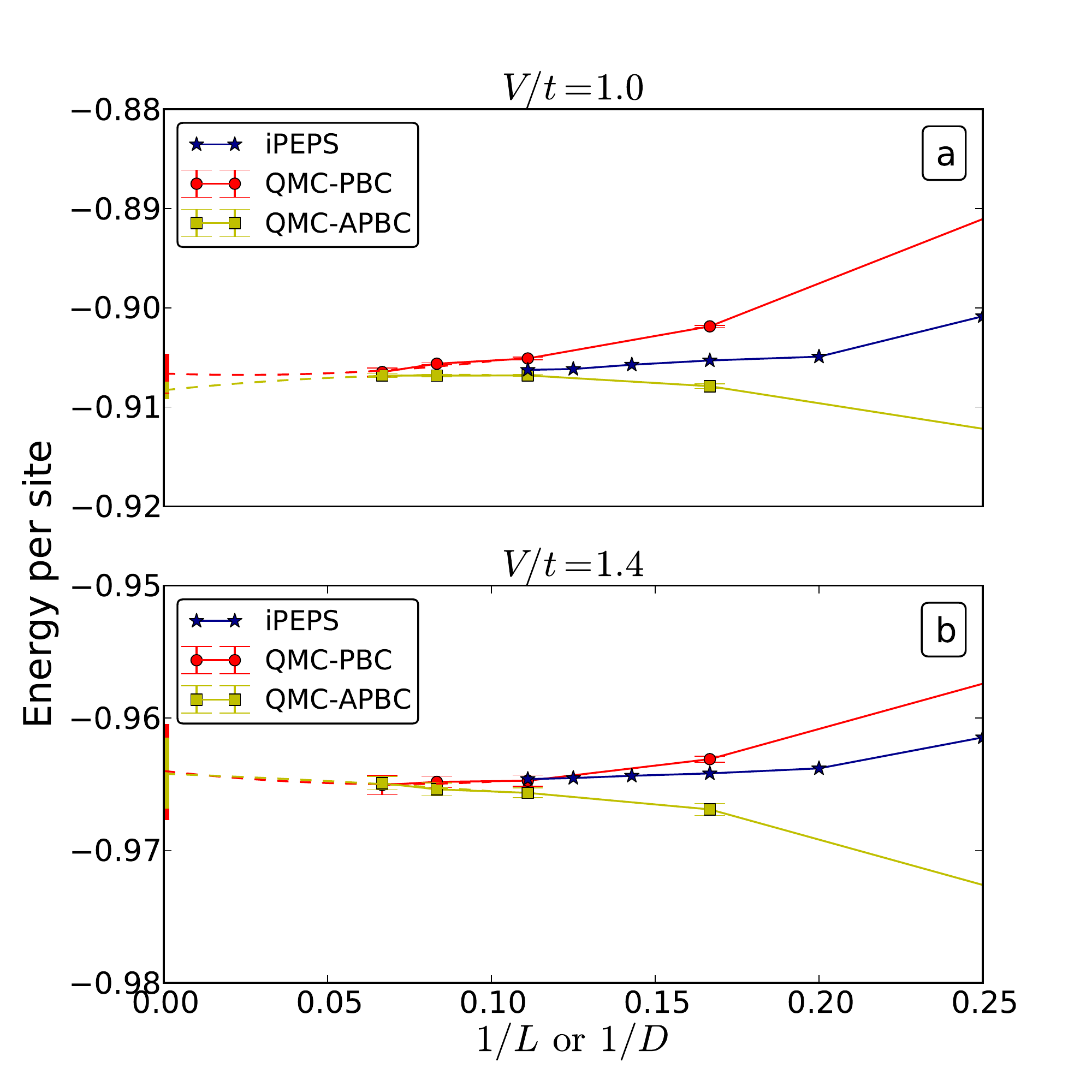}
\caption{Ground state energy per site of a honeycomb lattice versus inverse system length $1/L$ (QMC) or inverse bond dimension $1/D$ (iPEPS). The QMC results of periodic boundary conditions and anti-periodic boundary conditions approach to the thermodynamic limit value from different sides.}
\label{fig:energy}
\end{figure} 
 
\begin{figure}[t]
\includegraphics[width=9cm]{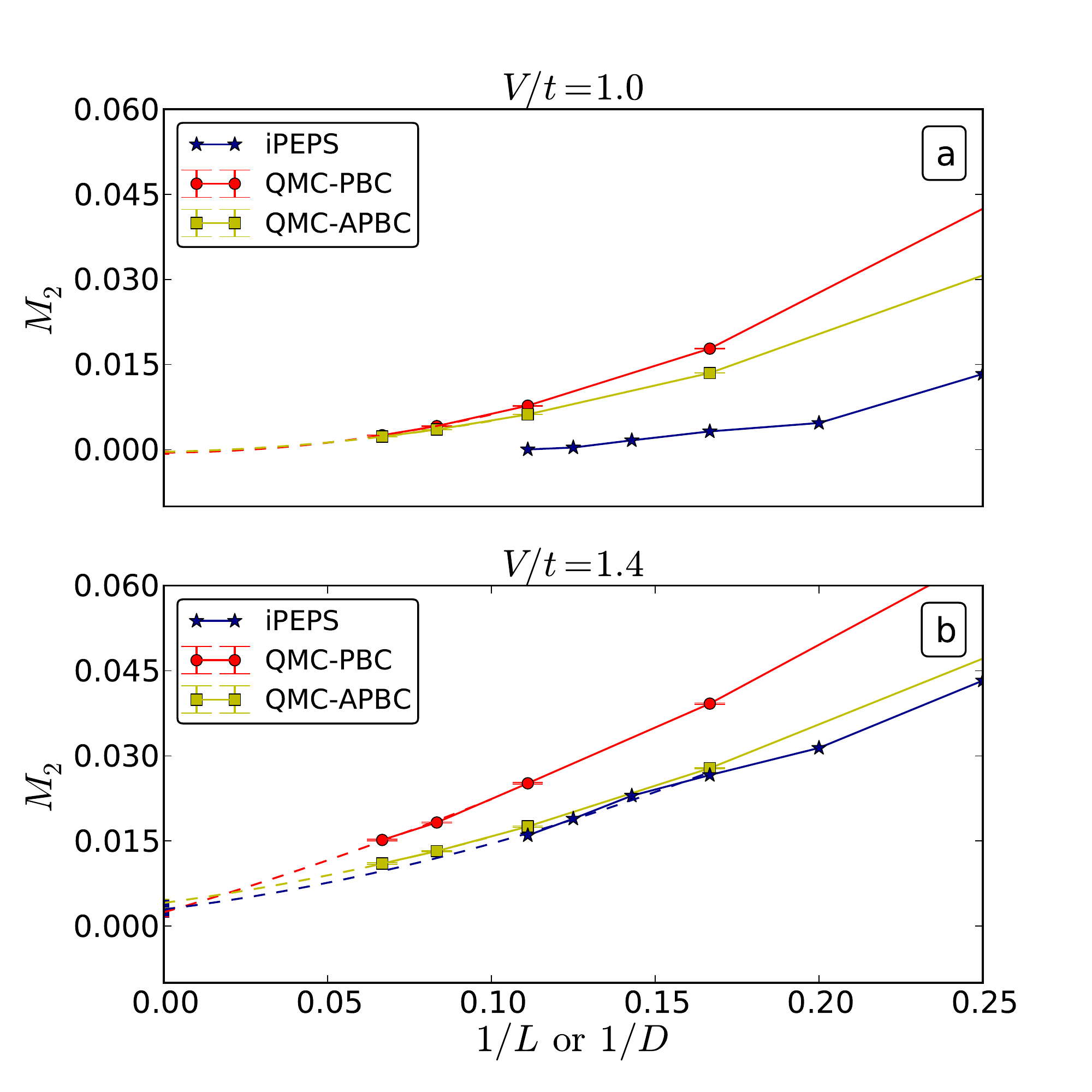}
\caption{The QMC results for the CDW structure factor compared with the square of CDW order parameter calculated using iPEPS. For $V/t=1$ the CDW order parameter vanishes for bond dimensions $D>8$ in iPEPS. For $V/t=1.4$ we used a linear fit in $1/D$ of the CDW order parameter to obtain an estimate in the infinite $D$ limit (see Ref.~\onlinecite{Wang:2014ib} for more details).}
\label{fig:cdw}
\end{figure}

Finally, we compare with infinite projected entangled-pair states (iPEPS) results obtained for the honeycomb lattice.~\cite{jordan2008,PhysRevB.81.165104, Wang:2014ib} iPEPS is a variational method which works in thermodynamic limit, whose accuracy can be systematically improved by increasing the bond dimension $D$. Figure~\ref{fig:energy} shows the ground state energy per site versus $1/L$ together with iPEPS results versus $1/D$.  QMC results for systems with periodic boundary conditions and those anti-periodic boundary condition along $x$-direction approach the $L\rightarrow\infty$ limit from different sides,  thus bracketing the ground state energy in the thermodynamic limit. Extrapolation of all data yields consistent results. Figure~\ref{fig:cdw} shows the CDW structure factor $M_{2}$ versus $1/L$, which extrapolates to the square of the CDW order parameter. iPEPS on the other hand can directly measure the order parameter since the symmetry is spontaneously broken for an infinite system. Extrapolation again yields consistent results and shows the system orders for $V/t=1.4$ but not at $V/t=1.0$.  

\begin{figure}[!t]
\includegraphics[width=9cm]{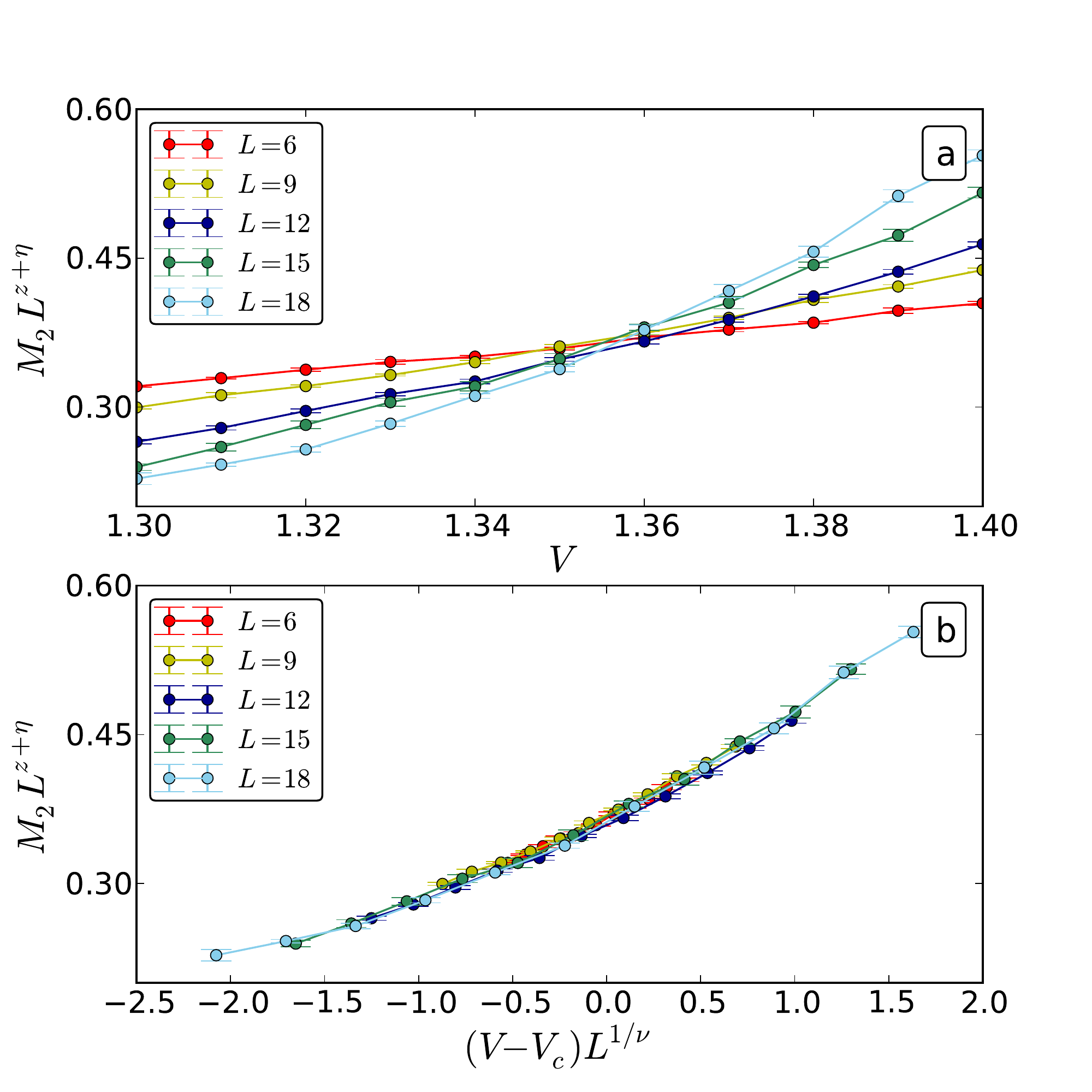}
\caption{(a) Scaled CDW structure factor of different system sizes cross at the transition point (b) Scaled CDW structures factor collapse on to a single curve when plotted against scaled interaction strength. }
\label{fig:scaleM2}
\end{figure}

\subsection{Fermionic Quantum Critical Point}
\label{sec:tvresults}
We finally apply the projector LCT-INT to study the quantum critical point of spinless $t$-$V$ model on a honeycomb lattice, which we previously studied by CT-INT simulations.\cite{Wang:2014ib}
Our calculations go beyond the previous results in two aspects. We can directly address the $T=0$ quantum critical point using the projection version of LCT-INT and we are able to reach larger system sizes up to $L=18$. Since a detailed finite size scaling study is beyond the scope of this paper, we use the critical values obtained in Ref.~\onlinecite{Wang:2014ib} and check for consistency. The CDW structure factor should follow the scaling ansatz 
\begin{equation}
M_{2} L^{z+\eta} = \mathcal{F}\left((V-V_{c}) L^{1/\nu}\right), 
\label{eq:FSS}
\end{equation}
where we previously found $z+\eta = 1.302$, $\nu = 0.8$ and $V_{c}/t=1.356$.
\cite{Wang:2014ib} Figure~\ref{fig:scaleM2}(a) shows the scaled CDW structure factor $M_{2} L^{z+\eta}$ where all curves cross around $V_{c}$ when using these critical exponents.  Scaling of the $x$-axis using $(V-V_{c}) L^{1/\nu}$ yields good data collapse, shown in Figure~\ref{fig:scaleM2}(b). We  conclude that the new zero temperature results on  larger system size are consistent with previous findings concerning critical point and critical exponents in Ref.~\onlinecite{Wang:2014ib}. 

\section{Discussion \label{sec:discussion}}
In this paper we presented details of the ground-state version of the LCT-INT algorithm of Ref.~\onlinecite{Mauro}. As a continuous-time QMC algorithm it eliminates the Trotter error due to time discretization of the BSS algorithm while still keeping the favorable \emph{linear scaling} with projection time and interacting strength. It is therefore well suited for simulations of the ground state of strongly correlated lattice fermions. 

Although the LCT-INT algorithms\cite{Mauro} and the projection version described here share operational similarities with the BSS algorithm,~\cite{Blankenbecler:1981vj, Sugiyama:1986vt, Sugiyama:1986vt, White:1989wh, Loh:1992, Assaad:2008hx} there are important differences. In the BSS formalism, one breaks the projection operator $e^{-\Theta \hat{H}}$ into small discrete time steps and performs Trotter-Suzuki decomposition for each time step, which leads to a systematic time-discretization error. 
The BSS algorithm then  decouples the interaction terms using auxiliary fields. A typical update scheme is to sweep through these time slices~\cite{Loh:1992, Assaad:2008hx} and flip the auxiliary fields, similar to our scheme of sweeping through the intervals. However, the time slices of the BSS algorithm are fixed in time and their number is proportional to the projection time. Each time slice contains $\mathcal{O}(N)$ auxiliary fields, therefore even with a brute force propagation of the Green's function on the site basis (Eq.~(\ref{eq:wrap})) one can achieve $\mathcal{O}(N^{3})$ scaling. While in our case the number and positions of vertices are allowed to fluctuate so we need to propagate in the eigenbasis (Eqs.~(\ref{eq:eigenbasisK}-\ref{eq:eigenbasisV})) and use $M\sim\langle k \rangle$ intervals such that on average each interval contains a single vertex to achieve a similar $\mathcal{O}(N^{3})$ scaling. 

Formally the Monte Carlo weight in Eq.~(\ref{eq:expansion}) is similar to the local weight of the CT-HYB method.~\cite{PhysRevLett.97.076405, Werner:2006iz} In particular the matrix version of CT-HYB\cite{Werner:2006iz} also evaluates the Monte Carlo weight in the eigenbasis of a propagator. However, our case is simpler because $e^{-\tau \hat{H}_{0}}$ is a single particle propagator and the Monte Carlo weight simplifies to a determinant instead of  than a trace in the many-body Hilbert space in CT-HYB. The present method can still benefit from algorithmic developments for the CT-HYB method. In particular a Krylov approach for imaginary time propagation~\cite{Lauchli:2009er} may  bring the cost of propagation of the Green's function to $\mathcal{O}(N^{2})$.
Our ``sweep through intervals'' scheme is also similar to the sliding window approach of the CT-HYB algorithm.~\cite{Shinaoka:2014dv} One may alternatively consider using trees or skip list data structure to store partial matrix products.~\cite{Gull:2011jd, PhysRevB.90.075149}

Besides being free of the discretization error, the continuous-time QMC approach provides a direct means to compute quantities such as observable derivatives Eq.~(\ref{eq:derv}), that are harder to obtain in discrete time simulations. These in turn may be used to locate interesting phase transitions with an accuracy that cannot easily be reached by standard discrete-time algorithms. Furthermore, the simple interaction dependency of the Monte Carlo weight $w(\mathcal{C})\sim V^{k}$ allows straightforward combination with the histogram reweighting~\cite{1988PhRvL..61.2635F, Ferrenberg:1989tf} or the Wang-Landau sampling~\cite{Wang:2001eba, Troyer:2003fta} techniques. Both approaches can produce results in a \emph{continuous range} of interaction strength by recording histograms over the perturbation order $k$. Combined with Eq.~(\ref{eq:derv}), the method offers new exciting opportunities to bring the study of quantum criticality of correlated fermions to a new level, approaching to what was achieved in the simulations of the classical~\cite{Ferrenberg:1991ta} and quantum spin systems.~\cite{Sandvik2010}


\section{Acknowledgments}
The authors thank Fahker Assaad, Jakub Imri\v{s}ka and Hiroshi Shinoka for helpful discussions. Simulations were performed on the M\"{o}nch cluster of Platform for Advanced Scientific Computing (PASC), the Brutus cluster at ETH Zurich and the ``Monte Rosa'' Cray XE6 at the Swiss National Supercomputing Centre (CSCS). We have used ALPS libraries~\cite{BBauer:2011tz} for Monte Carlo simulations and data analysis. DMRG results in Fig.~\ref{fig:nncorr} have been obtained using \verb|mps_optim| application~\cite{Dolfi:2014cc} of ALPS project. This work was supported by ERC Advanced Grant SIMCOFE, by the Swiss National Science Foundation through the National Competence Centers in Research QSIT and MARVEL, and the Delta-ITP consortium (a program of the Netherlands Organisation for Scientific Research (NWO) that is funded by the Dutch Ministry of Education, Culture and Science (OCW)).

\appendix 

\section{Site-Shift Update~\label{sec:siteshift}}
An optional additional update shifts a vertex between $X(i,j)$ by moving the  site $j$ to another neighbor of the site $i$ denoted by $j^{\prime}$. This will change the vertex matrix to $X(i, j^{\prime})$. It amounts to insert a vertex matrix $X(j, j^{\prime})$ at the same imaginary time without changing the perturbation order. The acceptance ratio is 
\begin{equation}
r_\mathrm{shift} = \frac{\det(LX(j, j^{\prime})R)}{\det(LR)} = -4 G_{jj^{\prime}} G_{j^{\prime}j}. 
\end{equation}
Since the site $j $ and $j^{\prime}$ belongs to the same sublattice, $G_{jj^{\prime}}  = -G_{j^{\prime}j}$ (see  
Eq.~(\ref{eq:symmetry})) ensures the acceptance ratio is positive. The formula for updating the  Green's function after a shift move is identical to Eq.~(\ref{eq:updateG}), with indices $i,j$ replaced by $j,j^{\prime}$.  

\section{Proof of Eq.~(\ref{eq:symmetry}) \label{sec:symmetry}}
Equation (\ref{eq:symmetry}) is easiest to prove in the finite temperature formalism. Suppose the trial wave function $|\Psi_{T}\rangle$ is the ground state of a noninteracting trial Hamiltonian $K_{T}$. The equal time Green's function can be formally written as 
\begin{equation}
G(\tau) = \lim_{\beta\to \infty}\left[\mathbb{I} + B(\tau, 0) e ^{-\beta K_{T}} B(\Theta, \tau) \right]^{-1}. 
\end{equation}
We introduce a diagonal matrix $D_{ij} =\eta_{i} \delta_{ij}$ and the bipartite conditions implies $D K D = -K$. Together with $X(i,j)^{-1} = X(i,j)$ this shows that $({B(\tau_{1}, \tau_{2})^{-1}})^{T}  = D B(\tau_{1}, \tau_{2}) D $. Similarly, assuming the trial Hamiltonian also satisfies $D K_{T} D = -K_{T}$, one has $e^{\beta K_{T}} = D e^{-\beta K_{T}} D $. Combing these facts it is then straightforward to show that 

\begin{eqnarray}
[G(\tau)]^{T} -\mathbb{I} & = & \lim_{\beta\to \infty} \left[\mathbb{I} + B(\Theta, \tau)^{T} e ^{-\beta K_{T}} B(\tau, 0)^{T} \right]^{-1} -\mathbb{I} \nonumber \\
& = & - \lim_{\beta\to \infty}\left[\mathbb{I} + ({B(\tau, 0) ^{-1}})^{T}e^{\beta K_{T}} ({B(\Theta, \tau)^{-1}})^{T} \right ] ^{-1} \nonumber \\
& = & - \lim_{\beta\to \infty}\left [\mathbb{I} + D B(\tau, 0)e^{-\beta K_{T}} B(\Theta, \tau)  D \right]^{-1} \nonumber \\
& = & - D G(\tau) D. 
\end{eqnarray}
This proves Eq.~(\ref{eq:symmetry}). 

\section{Proof of Eq.~(\ref{eq:updateG})  \label{sec:updateG}}

Notice that ${X}({i},{j})_{lm} = \delta_{lm}(1- 2\delta_{l{i}} -2 \delta_{m{j}}  )$, $G^{\pm}$  can be obtained from $G$ using the Sherman-Morrison formula twice:~\cite{Loh:1992, Assaad:2008hx}
\begin{eqnarray}
G^{\pm}_{lm} &= &G^{\prime}_{lm} + b G^{\prime}_{lj} (\delta_{jm} -G^{\prime}_{jm}) \label{eq:Gprime}, \\
G^{\prime}_{lm}  
&= &G_{lm} + a G_{li} (\delta_{im} -G_{im})  \label{eq:Gprimeprime}, 
\end{eqnarray}
where 
\begin{eqnarray}
a  &= & \frac{2 }{1 -2(1-G_{ii}) } \label{eq:a}, \\
b  &= & \frac{2 }{1 -2(1-G^{\prime}_{jj})} \label{eq:b}. 
\end{eqnarray}
Substituting Eqs.~(\ref{eq:Gprimeprime}-\ref{eq:b}) into Eq.~(\ref{eq:Gprime}) and using that for $i\neq j$,  $a^{-1} = 0$ and $ab = -1/({G_{ij} G_{ji}})$, one arrives at Eq.~(\ref{eq:updateG}). 
\bigskip

\section{Monte Carlo Updates in the Eigenbasis}
 \label{sec:eigen}
 
%
Once a Monte Carlo update is accepted, we update $\tilde{G} = U^{\dagger} G U$  using

\begin{eqnarray}
 \tilde{G}^{\pm}_{lm} = \tilde{G}_{lm}  &- &\frac{\left(\tilde{G}U^{\dagger}\right)_{l j} \left( U\tilde{G} - U \right)_{im} }{ (U\tilde{G}U^{\dagger})_{ij} }\nonumber \\ &-& \frac{\left(\tilde{G}U^{\dagger}\right)_{l i} \left( U\tilde{G}- U\right)_{jm}  }{ (U\tilde{G}U^{\dagger})_{ji} }
\end{eqnarray}
This update only involves matrix-vector multiplication and outer-product of column and row vectors, both requiring $\mathcal{O}(N^{2})$ operations.   

\bibliographystyle{apsrev4-1}
\bibliography{zeroT-CTBSS}


\end{document}